\documentclass[lettersize,journal]{IEEEtran}
\usepackage{amsmath,amsfonts}
\usepackage{algorithmic}
\usepackage{algorithm}
\usepackage{array}
\usepackage[caption=false,font=normalsize,labelfont=sf,textfont=sf]{subfig}
\usepackage{textcomp}
\usepackage{stfloats}
\usepackage{url}
\usepackage{cite}
\usepackage{verbatim}
\usepackage{graphicx}
\usepackage{float}
\usepackage{subfig}
\usepackage{booktabs}
\usepackage{multirow}
\usepackage{color}
\usepackage{makecell}
\usepackage{soul}
\usepackage[colorlinks,citecolor=blue,linkcolor=blue]{hyperref}

\usepackage[numbers,sort&compress]{natbib}

\makeatletter

\newcommand{\Rmnum}[1]{\expandafter\@slowromancap\romannumeral #1@}
\makeatother
\usepackage[justification=centering]{caption}
\def\BibTeX{{\rm B\kern-.05em{\sc i\kern-.025em b}\kern-.08em
    T\kern-.1667em\lower.7ex\hbox{E}\kern-.125emX}}
\usepackage{balance}
\begin{document}

\title{{Parameter Estimation of Mixed Gaussian-Impulsive Noise: An U-net++ Based Method}}

\author{Tianfu Qi, Jun Wang, \emph{Member, IEEE}, Xiaonan Chen, \emph{Student Member, IEEE}, Wei Huang
\thanks{Tianfu Qi is with the School of Information and Communication Engineering, University of Electronic Science and Technology of China, Chengdu 611731, China (e-mail: 2019270102009@uestc.edu.cn).}%
\thanks{Jun Wang, Xiaonan Chen and Wei Huang are with the National Key Laboratory of Science and Technology on Communications, University of Electronic Science and Technology of China, Chengdu 611731, China (e-mail: junwang@uestc.edu.cn; zslb200909@gmail.com; 18581810911@163.com.}\vspace{-0.2cm}
}

\markboth{Qi \MakeLowercase{\textit{et al.}}: Parameter Estimation of Mixed Gaussian-Impulsive Noise: An U-net++ Based Method}%
{Shell \MakeLowercase{\textit{et al.}}: A Sample Article Using IEEEtran.cls for IEEE Journals}
\maketitle

\begin{abstract}
In many scenarios, the communication system suffers from both Gaussian white noise and non-Gaussian impulsive noise. In order to design optimal signal detection method, it is necessary to estimate the parameters of mixed Gaussian-impulsive noise. Even though this issue can be well tackled with respect to pure mixed noise, it is quite challenging based on the received single-channel signal including both transmitting signal and mixed noise. To mitigate the negative impact of transmitting signal, we propose a parameter estimation method by utilizing a neural network, namely U-net++, to separate the mixed noise from the received single-channel signal. Compared with existing blind source separation based methods, simulation results show that our proposed method can obtain rather better performance in terms of estimation accuracy and robustness under various scenarios.
\end{abstract}



\section{Introduction}
\IEEEPARstart{M}{IXED} noise consisting of both non-Gaussian impulsive noise (IN) and Gaussian white noise (WGN) exists in many practical communication scenarios \cite{paper3}. Due to Brown movement of electron, the communication signal is inevitably influenced by WGN. Meanwhile, there are a lot of sources resulting in IN, such as lightning, underwater acoustic noise \cite{paper6}, broadband power lines \cite{paper2}, cosmic noise, etc. Even the duration of each impulse in IN is very short, it includes significant energy to distort the transmitting signal. In order to design optimal or near-optimal signal detection methods under mixed Gaussian-impulsive noise, it is necessary to estimate the parameters of the mixed noise.

The widely used IN model considering the physical generation mechanism is the symmetric $\alpha$ stable (S$\alpha$S) distribution \cite{paper6}. In fact, WGN is just a special case of S$\alpha$S distributed noise with $\alpha=2$. Even though closed-form probability density function (PDF) of S$\alpha$S distributed noise only exists for Gaussian and Cauchy ($\alpha=1$) noise, its characteristic function (CF) is quite simple and contains only two parameters \cite{paper10}. Many algorithms have been proposed to well estimate the parameters of S$\alpha$S distributed noise \cite{paper4,paper8,paper16}.

Compared with IN, it is rather difficult to model mixed Gaussian-impulsive noise with concise and closed-form PDF or CF. In \cite{paper3}, Sureka et al. raised a closed-form approximation to model the PDF of mixed Gaussian-impulsive noise, where S$\alpha$S distributed noise is considered. Their proposed PDF expression is determined by 5 parameters, and can match the practical PDF quite well. By utilizing pure noise samples, they also provided estimation methods of these parameters based on empirical characteristic function (ECF) presented in \cite{paper4}.

To well estimate the parameters of mixed noise, using pure noise samples is largely advocated. Consequently, communication transmitter has to keep silence for enough long time, which sacrifices spectral efficiency. Therefore, it is well motivated to estimate the mixed Gaussian-impulsive noise parameters based on the received signal including both transmitting signal and noise, so that the system overhead due to transmitter's silence can be avoided. However, the transmitting signal will seriously degrade the parameter estimation performance of mixed noise, especially  for the case that signal-to-noise ratio (SNR) is relatively large.

To mitigate the negative influence induced by the transmitting signal, it is necessary to first separate mixed noise from the received signal. Essentially, this task can be considered as a problem of blind source separation (BSS). If the signal samples can be received through multiple channels, e.g. at a multi-antenna receiver, the corresponding BSS problem could be well handled \cite{book1,paper1}. However, in some practical scenarios, the signal samples can only be gathered through single-channel receiver due to the constraint of size, weight and power. For this case, it is rather challenging to separate the mixed noise from transmitting signal with existing statistical signal processing based methods.

In recent years, the fast advances of deep learning provide new possible solution for single-channel BSS problem. Ronneberger first proposed the a neural network called after U-net, and employed it in medical image segmentation \cite{paper9}. U-net is a kind of fully convolutional network and contains two main parts: encoder sub-network and decoder sub-network to get context information and localization, respectively. Siddiquee et al. further built an improved U-net named after U-net++ by introducing a supervision mechanism and improved skip connections to enhance the network learning ability \cite{paper14}. In addition, gradient descent is performed with respect to the sum of the losses of multiple convolutional layers, so that some potential features can be remained and output during the training process as the neural network becomes deeper.

In this letter, we consider the parameter estimation of mixed Gaussian-impulsive noise based on received single-channel signal including both transmitting signal and noise. To circumvent the difficulty in single-channel BSS and inspired by the potential of U-net++, we propose an U-net++ based method to separate mixed noise from received signal. We first introduce a preprocessing module before the U-net++ to mitigate the negative effect of impulsive noise and improve robustness. Then, an elaborately designed U-net++ is used to separate the transmitting signal from received signal. Finally, the mixed noise parameter can be estimated by utilizing ECF based method. Compared with existing BSS methods, simulation results show remarkable performance gain of our proposed algorithm in various scenarios.

\vspace{-0.2cm}

\section{Model and Problem Description}
In this section, we describe the signal model and the corresponding parameter estimation problem.

\vspace{-0.2cm}

\subsection{Signal Model}
The received signal samples can be represented as follows:
\vspace{-0.2cm}
\begin{equation}
y_i=s_i+ \underbrace{ n_{gi}+n_{si} }_{n_{mi}} ,i=1,2,...,N
\vspace{-0.1cm}
\label{ReceivedSignal}
\end{equation}
where $s_i$ represents transmitting signal sample, which can be generated with popular modulation scheme. $n_{gi}$ and $n_{si}$ are independent and identically distributed (i.i.d) WGN and IN samples, respectively. In this paper, IN sample $n_{si}$ is modeled to follow the S$\alpha$S distribution with characteristic exponent $\alpha$ and scale parameter $\gamma_s$. Meanwhile, WGN sample $n_{gi}\sim \mathcal{N}(0,\sigma^2) $  can be expressed as S$\alpha$S model with $\alpha =2$ and  $\sigma^2=2\gamma_g^2$ \cite{paper16}. By assuming that WGN and IN samples are mutually independent, a closed-form  and well approximate PDF is proposed for mixed noise in \cite{paper3}. This model is mainly controlled by parameters $\alpha$, $\gamma_s$ and $\gamma_g$. In this paper, we try to estimate these parameters so that signal detection algorithm can be further designed.

To describe diverse mixed noise scenarios, we define $\lambda$ as the strength ratio between WGN and IN, which is given as follows,
\vspace{-0.1cm}
\begin{equation}
\lambda=\frac{\gamma_g^2}{\gamma_s^{\alpha}}
\end{equation}
Larger $\lambda$ implies that WGN takes greater proportion in the mixed noise. When $\lambda$ approaches +$\infty$ or 0, mixed noise will degenerate to WGN or IN, respectively.

\begin{figure*}[htbp]
\centering
\includegraphics[width=6.5in]{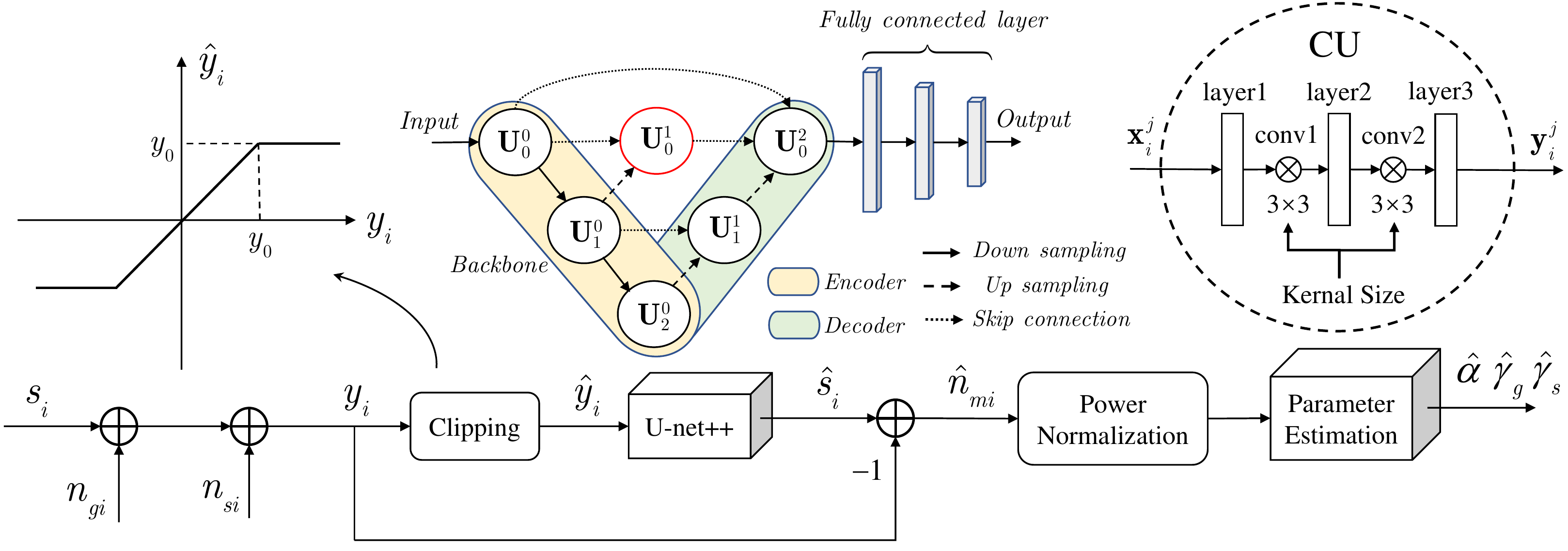}
\caption{Algorithm framework}
\label{fig_2}
\end{figure*}

\vspace{-0.3cm}

\subsection{Problem Description}
In this paper, we estimate the parameters $\alpha$, $\gamma_s$ and $\gamma_g$ of mixed noise based on the received signal samples $y_i$, $i=1,2,\cdots,N$. Then, we can further get the parameter $\lambda$.

To mitigate the negative effect of transmitting signal sample $s_i$, the key is to separate the mixed noise sample $n_{mi}$ from received signal $y_i$. Equivalently, the approximate `pure' mixed noise samples can be obtained by cancelling the transmitting signal as following:
\vspace{-0.1cm}
\begin{equation}
\hat{n}_{mi}=y_i-\hat{s}_i, i=1,\cdots,N \label{mixed_noise}
\vspace{-0.1cm}
\end{equation}
where $\hat{s}_i$ is the separated version of transmitting signal $s_i$. The optimal maximum likelihood estimation (MLE) of  ${s}_i$ can be basically expressed as following:
\vspace{-0.2cm}
\begin{equation}
\hat{s}_i=\mathop{\arg\max}\limits_{s_i\in\Omega}\log\left[ g(y_i-s_i)\right] \label{ML}
\vspace{-0.1cm}
\end{equation}
where $g(\cdot)$ is the PDF of mixed noise, $\Omega$ represents the adopted modulation alphabet. As mentioned before, the PDF of mixed noise is determined by the parameters $\alpha$, $\gamma_s$ and $\gamma_g$ to be estimated. One possible strategy is to first estimate the parameters of mixed noise based on $y_i,i=1,\cdots,N$ by omitting the transmitting signal $s_i$. Then, the transmitting signal can be estimated and cancelled based on \eqref{ML} and \eqref{mixed_noise}. By alternatively performing these two operations, the estimation performance may be expected to be improved iteratively. However, our simulation results in Section \ref{simulation} reveal that this strategy can not work well, especially for the case with relatively large SNR. In next Section, we propose a novel neural network based method to separate mixed noise and transmitting signal.

\vspace{-0.3cm}
\section{U-net++ Based Approach}
\vspace{-0.1cm}
In this section, we describe the proposed parameters estimation framework and the adopted signal separation network U-net++.
\vspace{-0.4cm}

\subsection{Algorithm Framework}
\vspace{-0.1cm}
Our proposed mixed noise parameter estimation algorithm framework is illustrated in Fig.\ref{fig_2}. The received signal $y_i$ given in \eqref{ReceivedSignal} is first clipped with threshold $y_0$ to alleviate the influence of IN component $n_{si}$. We will discuss the choice of $y_0$ in what follows. Then, the clipped version $\hat{y}_i$ is fed into the neural network U-net++ to separate the transmitting signal from mixed noise. By cancelling the separated transmitting signal $\hat{s}_i$ based on \eqref{mixed_noise}, we get the approximate `pure' mixed noise $\hat{n}_{mi}$. Subsequently, power normalization is performed with respect to $\hat{n}_{mi}$ to enhance the algorithm stability, the mixed noise parameters are then estimated by utilizing ECF based method to get $\hat{\alpha}$, $\hat{\gamma}_g$, and $\hat{\gamma}_s$.

\vspace{-0.4cm}

\subsection{Clipping}
\vspace{-0.1cm}
Clipping is first performed to remove the outliers in received signal and avoid network overfitting. The clipping threshold $y_0$ can be determined according to Pauta criterion \cite{book2}. If the standard deviation of WGN is $\sigma$, it is almost impossible that the amplitude of WGN sample is larger than $4\sigma$. Therefore, $y_0$ could be $4\sqrt{2}\gamma_g+\max(s_i)$. $\max(s_i)$ is determined by transmitter modulation scheme, and can be consider as priori information. Without loss of generality, we set $\max(s_i)=1$ in this paper. However, as $\gamma_g$ is unknown for the receiver, it has to be set empirically.

Nikias has proven that the $\mathnormal{p}$-th moment of S$\alpha$S distributed noise exists when $-1<p<\alpha$ \cite{paper16}. Therefore, we have
\begin{equation}
\begin{split}
E[|n_{mi}|^p]&=E[|n_{gi}|^p]+E[|n_{si}|^p]\\
&=C_{p,2}\gamma_g+C_{p,\alpha}\gamma_s \label{moment}
\end{split}
\end{equation}
and,
\begin{equation}
C_{p,\alpha}=\frac{2^{p+1}\Gamma\left(\frac{p+1}{2} \right)\Gamma\left(\frac{-p}{\alpha} \right)}{\alpha \sqrt{\pi}\Gamma\left(\frac{-p}{2} \right)}
\vspace{-0.1cm}
\label{coefficient}
\end{equation}
where Gamma function $\Gamma(z)=\int_{0}^{\infty}t^{z-1}e^{-t}dt$. This model can approximate mixed noise well when $\alpha \in (0.5,1.9)$ \cite{paper3}. The right hand side (RHS) of (\ref{moment}) will approach infinite when $\alpha$ converges to 0. Thus, we consider $\alpha\in(0.5,1.9)$ in this paper. Note that the mixed noise can be considered to be approximate WGN for $\alpha>1.9$. To empirically determine threshold $y_0$ and ensure that $E(|n_{mi}|^p)$ always exists, we set $p=0.5$ and $\gamma_g:\gamma_s=1:1$. Meanwhile, we use $C_{0.5,0.51}$ to substitute $C_{0.5,0.5}$ in \eqref{coefficient}, which is not defined in complex domain. To facilitate calculating, we further replace $E[\sqrt{|n_{mi}|}]$ by $E[\sqrt{|y_{i}|}] $ and set threshold $y_0=4\sqrt{2} E[\sqrt{|y_{i}|}]/(C_{0.5,2}+C_{0.5,0.51})+1$. The simulation results in Section \ref{simulation} show that this threshold works well in various scenarios.

\vspace{-0.4cm}

\subsection{U-net++ Based Separation Network}
\vspace{-0.1cm}
U-net++ is a symmetric neural network consisting of encoder sub-network and decoder sub-network, which can extract the characteristics of data and reconstruct information, respectively \cite{paper14}. As a variant of U-net, modified skip pathway and deep supervision are introduced into U-net++ architecture. The skip pathway is composed of dense convolution blocks. It can bridge semantic gap between feature maps of encoder sub-network and decoder sub-network. U-net++ has been proved to be rather effective in recovering fine-grained details of the target objects \cite{paper14}. Meanwhile, U-net++ has the same dimension for input and output. In this paper, we utilize these merits of U-net++ to separate transmitting signal $s_i$ from single-channel signal $\hat{y}_i$, so that the distortion between the obtained $\hat{s}_i$  and $s_i$ can be as little as possible.

The adopted U-net++ is illustrated in Fig.\ref{fig_2}. Each circle in U-net++ network represents a convolution unit (CU), which is composed of three convolutional layers with the same kernel size. Let $\mathbf{U}_i^j$ be the $j$th CU in layer $i$. $\mathbf{x}_i^j$ and $\mathbf{y}^j_i$ denote the input and output of $\mathbf{U}_i^j$, respectively. In Fig.\ref{fig_2}, solid line, dashed line and dotted line represent down-sampling, up-sampling and concatenation operation, respectively. They can be mathematically expressed as follows,
\vspace{-0.1cm}
\begin{gather}
\vspace{-0.1cm}
\mathcal{D}(\mathbf{x})=[\max(x_1,x_2),\cdots,\max(x_{N-1},x_N)]\\
\mathcal{U}(\mathbf{x})=\left( x_1,x_1,x_2,x_2,\cdots,x_N,x_N\right)\\
\mathcal{C}(\mathbf{x}_1,\mathbf{x}_2)=( x_{1_1},x_{1_2},\cdots,x_{1_N},x_{2_1},\cdots,x_{2_N})
\end{gather}
where $\mathbf{x}=(x_1,x_2,\cdots,x_N)$, $\mathbf{x}_1=(x_{1_1},x_{1_2},\cdots,x_{1_N})$ and $\mathbf{x}_2=(x_{2_1},x_{2_2},\cdots,x_{2_N})$. The architecture of U-net++ is basically determined by the parameters of backbone, including the network depth and the channels of each CU. These parameters have to be determined based on the tradeoff between separation performance and complexity. Usually, as the depth of backbone becomes deeper, the time consumption will increase. Meanwhile, the dimension of the first fully connected layer will become larger due to the concatenation operation. Moreover, it may result in underfitting due to significant difference of dimensions between adjacent layers. Therefore, we set dropout layer in CU, and enough fully connected layers are added to ensure smooth propagation of errors.

As there may still exist some samples with relatively large magnitude despite the clipping, leaky relu is chosen as the activation function to prevent gradient vanishing or exploding. In this paper, normalized mean square error (NMSE) is chosen as the loss function because the manipulations are with respect to the signal waveform. The loss function $\mathcal{L}(\Theta)$ with network parameters $\Theta$ is given as follows,
\begin{equation}
\mathcal{L}(\Theta)=\frac{1}{N}\sum_{i=1}^{N}|\mathcal{F}(\hat{y}_i;\Theta)-\hat{s}_i|^2+\mathcal{R}(\Theta)
\end{equation}
where $\mathcal{F}(\hat{y}_i;\Theta)$ represents the output of network. The regularization item $\mathcal{R}(\Theta)$ is used to avoid overfitting. It is set as $\mathcal{R}(\Theta)=\mu||\Theta||_2$ with weight factor $\mu$.

\vspace{-0.1cm}
\subsection{Parameter Estimation}
Finally, we utilize ECF-based algorithm presented in \cite{paper4} to estimate mixed noise parameters $\alpha$, $\gamma_g$ and $\gamma_s$ based on $\hat{n}_{mi}$. As too large or small values of noise samples will lead to significant difference between CF and ECF, and hence affect algorithm stability, parameter estimation is performed with respect to $\hat{n}_{mi}$ which is the power normalized version of $n_{mi}$.

\vspace{-0.1cm}
\subsection{Complexity}
The complexity mainly comes from U-net++ and ECF based parameters estimation. The calculation burden of U-net++ is due to the convolution operation. We denote the time consumption of each floating-point operation to be $T$. Let the backbone depth of U-net++ and the size of kernel be $N$ and $M$, respectively. We denote the channel number of kernel in the $i$-th layer to be $C_i,i=0,\cdots,N$ and $C_0=1$. Network will be trained for $E$ epoches. Therefore, the complexity of U-net++ will be $\mathcal{O}\left[ EM^2 T\sum_{i=0}^{N-1}(N-i)C_iC_{i+1} \right]$. For parameter estimation, we utilize binary search to figure out the approximate solution of the transcendental equation of $\alpha$. Let the number of subdivision points be $N_b$, and we will calculate $E_p$ times to increase accuracy of parameter estimation. Finally, the overall complexity of algorithm can be simplified as $\mathcal{O}(EM^2N^2C_mT+E_pN_b\log N_b)$ where $C_m=\max\{C_iC_{i+1}\},i=0,\cdots,N-1$.


\section{Simulations}\label{simulation}
\vspace{-0.1cm}
In this section, we verify the performance of the proposed algorithm through simulations.

\vspace{-0.3cm}
\subsection{Training Dataset and Network Configuration}
\vspace{-0.1cm}
To obtain training dataset, we generate $s_i$  based on several modulation schemes including MSK, QPSK and 16-QAM. Mixed noise sample $n_{mi}$ is generated as the sum of i.i.d. WGN sample $n_{gi}$ and IN sample $n_{si}$ with different parameters $\alpha$ and $\lambda$ to represent different noise scenarios. To decrease truncation error, we set the dynamic range of IN sample $P\left[ |n_{si}| \leq R \right]>0.999$, and the step to be 0.01. In this paper, we adopt the algorithm proposed in \cite{paper20} to stably generate IN samples due to the absence of closed-form PDF expression for S$\alpha$S distribution. Note that S$\alpha$S distributed noise has no limited second moment except $\alpha\neq 2$. To measure the relationship between signal and noise, we adopt the metric called general signal-to-noise ratio (GSNR) as $GSNR(\mathrm{dB})=10\log_{10}\left[E_s/2\left(\gamma_s^{\alpha}+\gamma_g^2 \right) \right]$ with $E_s=\mathbb{E}[|{s_i}|^2]$.

For network training, the whole data is divided into mini-batch with size 200. The data corresponding to different parameter pair $(\alpha,\lambda,GSNR)$ is randomly selected to train the network with epoch 30. The parameter sets to generated the training dataset are $\alpha=\{1.0,1.2,1.4,1.6,1.8\}$, $\lambda=\{0.2,1,5\}$ and $GSNR=\{0,10,20\}$. The adopted hyper-parameters of U-net++ is given in Table \ref{ConfigUnet++}.

\linespread{1.0}
\begin{table}[htbp]
\begin{center}
\caption{Configuration of U-net++}\label{ConfigUnet++}
\scalebox{0.9}{
\begin{tabular}{cc}
\toprule
Parameter& Value\\
\midrule
\makecell[c]{Channels of CU \\of backbone}& $\textbf{U}_0^0,\cdots, \textbf{U}_5^0=\{16,32,60,96,144,256\}$ \\
Padding mode& `Same'\\
Kernel size& $3\times3$\\
Optimizer& \makecell[c]{`Adam' with $lr=0.001$, \\$\beta_1=0.9$, $\beta_2=0.999$\\ and $\epsilon=1e-8$}\\
Pooling mode& `Max-pooling'\\
Dropout rate& 0.3\\
Activation function& \makecell[c]{leaky relu with\\ $slope=0.01$} \\
\bottomrule
\end{tabular}
}
\end{center}
\end{table}
\linespread{1}

\vspace{-0.5cm}

\linespread{1.0}
\begin{table}[htbp]
\begin{center}
\caption{Estimation performance of $\alpha$}
\scalebox{0.8}{
\begin{tabular}{ccccccc}
\toprule
True $\alpha$& $\lambda$& GSNR&None& MSK& QPSK& 16-QAM\\
\midrule
 \multirow{9}{*}{1.2}&\multirow{3}{*}{0.1} & 0&1.183&1.211 &1.193 &1.204\\
 & & 10&1.192& 1.219& 1.213&1.223\\
 & & 20&1.178& 1.209& 1.166&1.192\\
  \cline{2-7}
 &\multirow{3}{*}{1} & 0&1.181& 1.175& 1.168&1.159\\
 & & 10&1.192& 1.200& 1.222&1.230\\
 & & 20&1.134& 1.191& 1.231&1.211\\
  \cline{2-7}
 &\multirow{3}{*}{10} & 0&1.190& 1.183& 1.184&1.226\\
 & & 10&1.181& 1.181& 1.176&1.192\\
 & & 20&0.896& 1.179& 1.181&1.165\\
  \cline{1-7}
 \multirow{9}{*}{1.8}&\multirow{3}{*}{0.1} & 0&1.825& 1.790& 1.765&1.758\\
 & & 10&1.726& 1.804& 1.790&1.802\\
 & & 20&1.344& 1.787& 1.796&1.778\\
  \cline{2-7}
 &\multirow{3}{*}{1} & 0&1.812& 1.793& 1.789&1.749\\
 & & 10&1.609& 1.798& 1.769&1.813\\
 & & 20&0.143& 1.779& 1.788&1.798\\
  \cline{2-7}
 &\multirow{3}{*}{10} & 0&1.766& 1.842& 1.873&1.865\\
 & & 10&1.184& 1.833& 1.813&1.853\\
 & & 20&0& 1.786& 1.789&1.844\\
\bottomrule \label{alpha}
\end{tabular}
}
\end{center}
\end{table}
\linespread{1}

\vspace{-0.5cm}
\subsection{Estimation Performance}
\vspace{-0.1cm}
Under different noise scenarios and modulation schemes, the estimation performances of $\alpha$, $\gamma_g$ and $\gamma_s$ are presented in Table \ref{alpha} and \ref{gamma}, respectively. These results are obtained as the average of 100 rounds of estimation. The results presented in the column `None' means that the parameters are directly estimated based on received single-channel signal $y_i$. We let it be the baseline. It can be seen that our proposed method can effectively estimate $\alpha$, $\gamma_g$ and $\gamma_s$ under diverse scenarios with different GSNR, $\lambda$ and modulation schemes of transmitting signal.

\linespread{1.1}
\begin{table}[htbp]
\begin{center}
\caption{Estimation performance of $\gamma_g$ and $\gamma_s$}
\scalebox{0.7}{
\begin{tabular}{ccccccc}
\toprule
$\alpha$& $\lambda (\gamma_g,\gamma_s)$& GSNR&None& MSK& QPSK& 16-QAM\\
\midrule
 \multirow{9}{*}{1.2}&0.1(0.477,1.982)& 0& (0.534,1.957)&(0.447,2.031) &(0.507,1.946) &(0.404,1.999)\\
 &0.1(0.151,0.291) & 10& (0.154,0.287)& (0.183,0.302)& (0.177,0.307)&(0.176,0.294)\\
 &0.1(0.048,0.043) & 20& (0.051,0.042)& (0.057,0.038)& (0.051,0.037)&(0.047,0.039)\\
  \cline{2-7}
 &1(1.118,1.204) & 0& (1.137,1.178)&(1.173,1.136)& (1.179,1.144)&(1.209,1.189)\\
 &1(0.354,0.177) & 10& (0.358,0.172)&(0.352,0.173)& (0.335,0.192)&(0.355,0.180)\\
 &1(0.112,0.026) & 20& (0.121,0.018)&(0.113,0.025)& (0.115,0.023)&(0.109,0.028)\\
  \cline{2-7}
 &10(1.508,0.291) & 0& (1.501,0.283)&(1.512,0.283)& (1.490,0.314)&(1.498,0.313)\\
 &10(0.477,0.043) & 10& (0.538,0.018)&(0.479,0.039)& (0.479,0.038)&(0.477,0.042)\\
 &10(0.151,0.006) & 20& (0.179,0.001)&(0.152,0.005)& (0.151,0.005)&(0.152,0.005)\\
  \cline{1-7}
 \multirow{9}{*}{1.8}&0.1(0.477,1.578) & 0& (0.299,1.621)&(0.509,1.525)& (0.421,1.411)&(0.466,1.381)\\
 &0.1(0.151,0.439) & 10& (0.297,0.352)&(0.134,0.444)& (0.183,0.407)&(0.119,0.441)\\
 &0.1(0.048,0.122) & 20& (0.486,0.066)&(0.063,0.114)& (0.058,0.118)&(0.070,0.108)\\
  \cline{2-7}
 &1(1.118,1.132) & 0& (1.098,1.152)&(1.144,1.102)& (1.132,1.116)&(1.161,1.052)\\
 &1(0.354,0.315) & 10& (0.445,0.171)&(0.352,0.322)& (0.381,0.279)&(0.335,0.333)\\
 &1(0.112,0.088) & 20& (0.663,0)&(0.118,0.077)& (0.113,0.082)&(0.111,0.084)\\
  \cline{2-7}
 &10(1.508,0.439) & 0& (1.516,0.408)&(1.479,0.429)& (1.449,0.409)&(1.443,0.478)\\
 &10(0.477,0.122) & 10& (0.499,0.008)&(0.471,0.130)& (0.475,0.127)&(0.463,0.154)\\
 &10(0.151,0.034) & 20& (0.759,0)&(0.151,0.029)& (0.152,0.029)&(0.149,0.039)\\
\bottomrule \label{gamma}
\end{tabular}
}
\end{center}
\end{table}
\linespread{1}

\begin{figure}[htbp]
\centering
\subfloat[$\alpha=1.0$]{\includegraphics[width=1.65in]{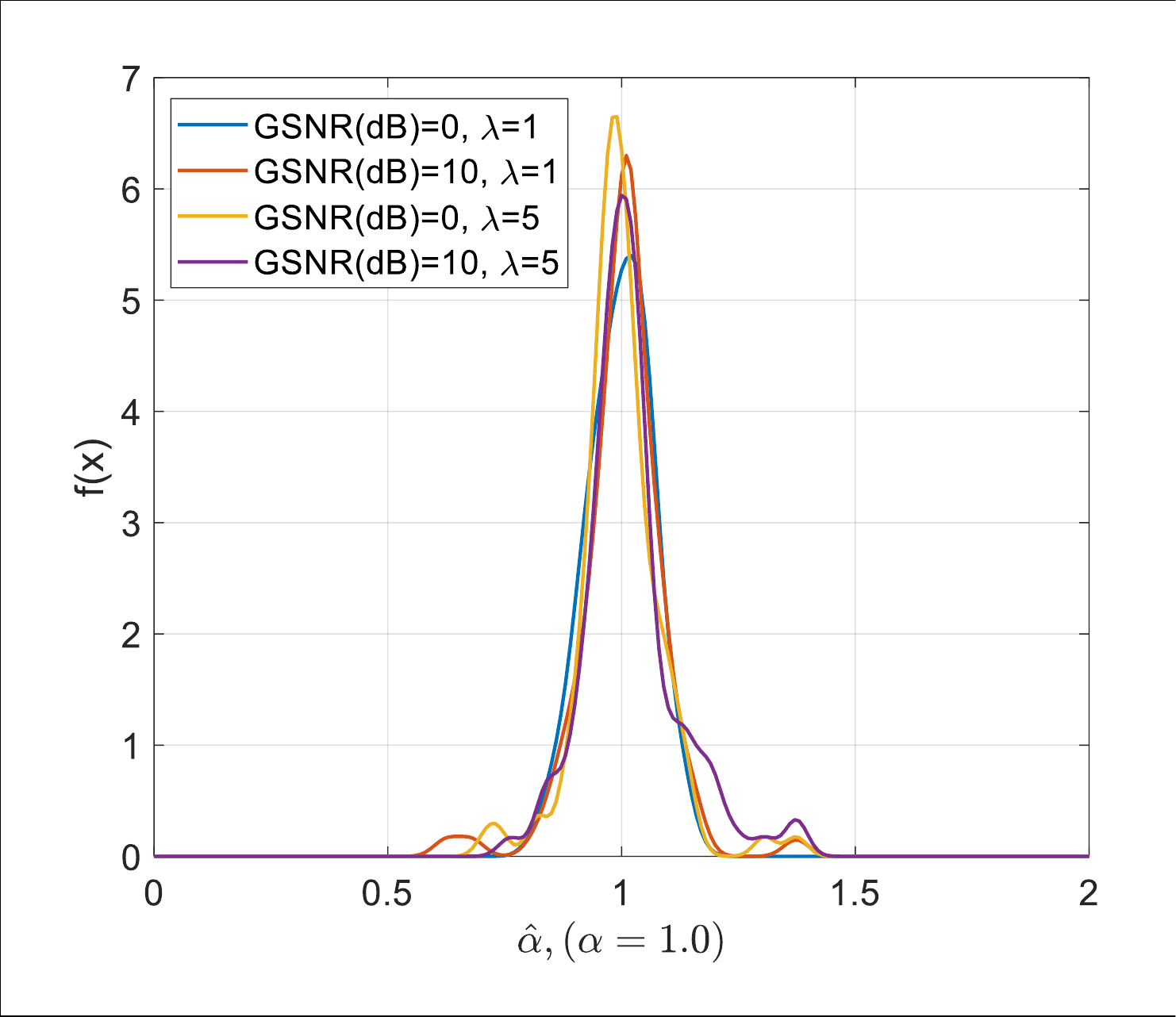}}
\subfloat[$\alpha=1.2$]{\includegraphics[width=1.65in]{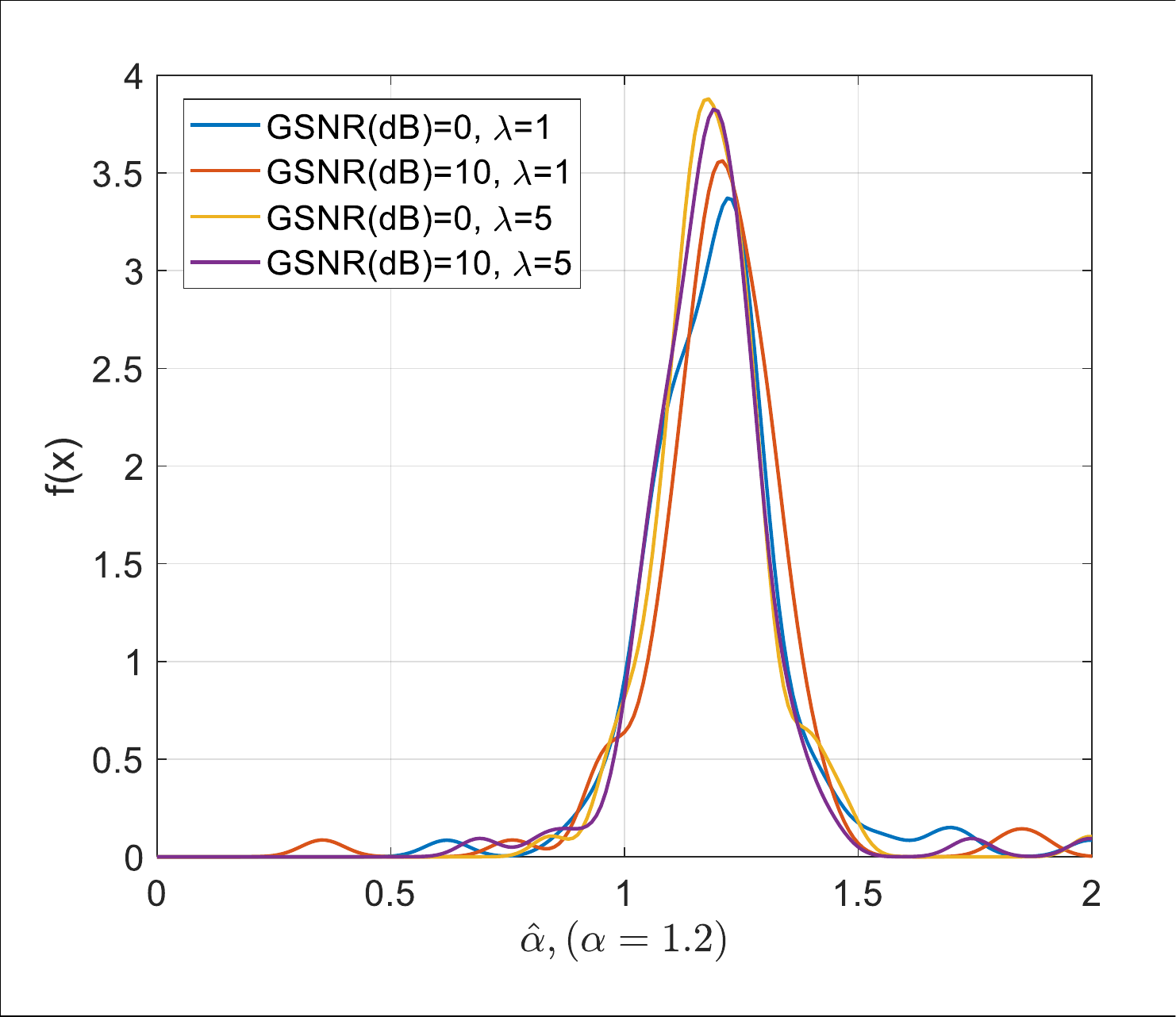}}

\subfloat[$\alpha=1.5$]{\includegraphics[width=1.65in]{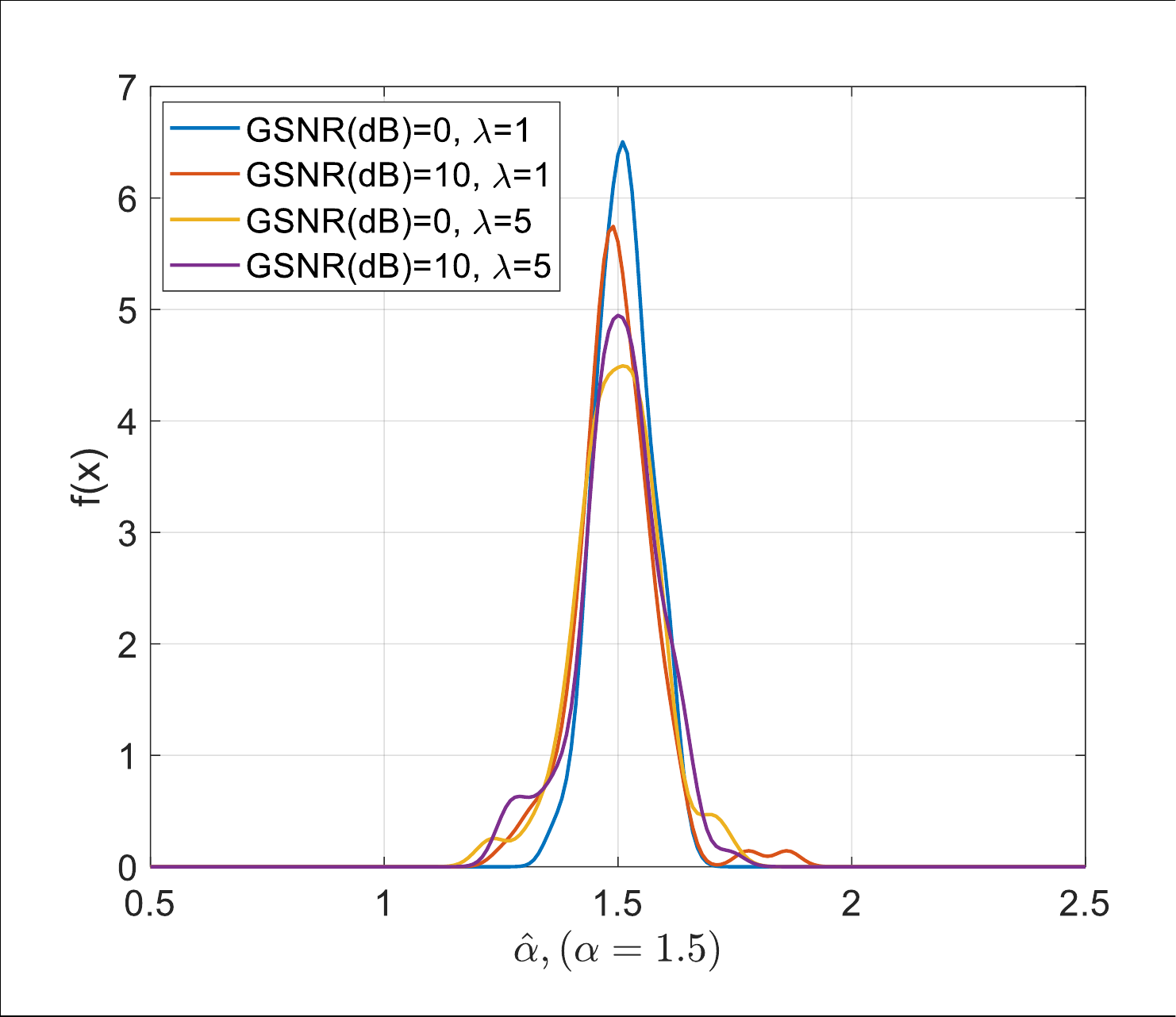}}
\subfloat[$\alpha=1.8$]{\includegraphics[width=1.65in]{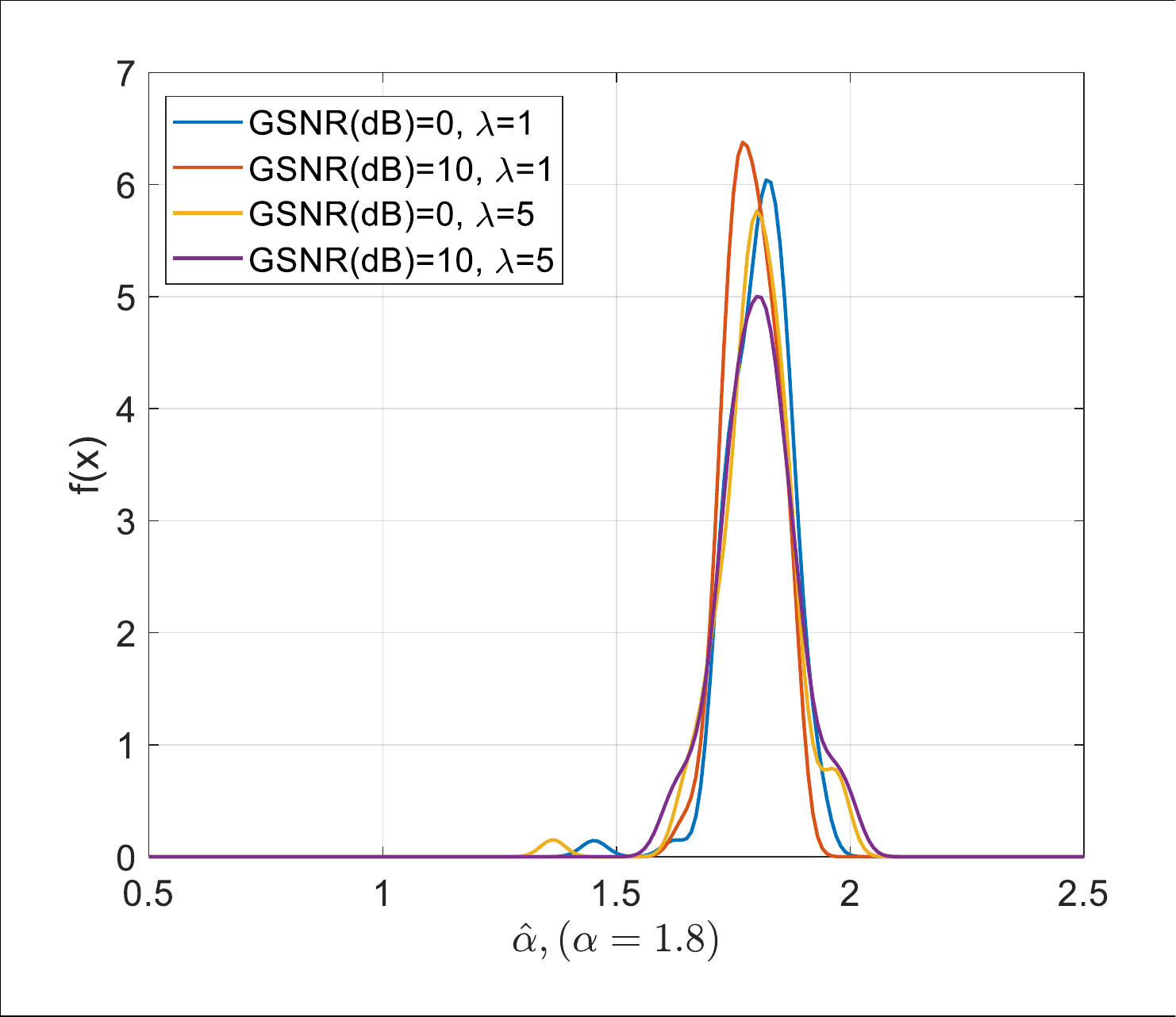}}
\caption{Estimation results distribution for different $\alpha$}
\label{fig_3}
\end{figure}

The estimated distribution of $\alpha$ is further presented in Fig.\ref{fig_3}. Due to limited space, we omit the estimation results distributions of $\gamma_g$ and $\gamma_s$ as the similar results are observed. It can be found that the variance of estimation results will become larger for the cases with relative large $\alpha$, GSNR and $\lambda$. The underlaying reason is that it is difficult for the network to learn the characteristics of IN as the IN component in mixed noise becomes smaller.

\vspace{-0.3cm}

\vspace{-0.2cm}
\subsection{Estimation Performance Comparison of Other Algorithms}
To validate the advantage of our algorithm, we further compare it with some other BSS based baselines. The following baselines are considered:

\begin{enumerate}
\item{Spectral Subtraction (SS)} \cite{paper17}: The transmitting signal is obtained by subtracting the spectrum of noise from the spectrum of received signal.
\item{Clipping} \cite{paper6}: All outliers, whose amplitudes are greater or equal to a specific threshold, are set to the given threshold.
\item{Filter based methods} \cite{paper15}: The noise can be suppressed adaptively by a filter, such as Least Mean $\mathnormal{p}$-norm (LMP) or Kalman filter.
\item{Empirical Mode Decomposition (EMD)} \cite{paper7}: The signal is decomposed into some oscillatory components called intrinsic mode function (IMF) to suppress the noise.
\end{enumerate}

Due to space limitations, we only present the performance comparison with respect to $\alpha$. Similar results have been observed for $\gamma_g$ and $\gamma_s$. The estimation performance of $\alpha$ is compared in Table \ref{methods}. It is noteworthy that the parameters of baselines are elaborately designed to get desirable performance. The running time comparison of these algorithms are given in Table \ref{time}.

\vspace{-0.2cm}
\linespread{1.1}
\begin{table}[htbp]
\begin{center}
\caption{Performance comparison of different algorithms}
\scalebox{0.8}{
\begin{tabular}{cccccccccc}
\toprule
$\alpha$& $\lambda$& GSNR& None& U-net++&SS&Kalman&EMD&LMP&Clipping\\
\midrule
 \multirow{9}{*}{1.2}&\multirow{3}{*}{0.1} & 0& 1.193& \textbf{1.211}& 1.205& 1.196& 1.195& 1.198&1.601\\
 & & 10& 1.195& \textbf{1.219}& 1.203& 1.204& 1.202& 1.199&1.392\\
 & & 20& 1.197& \textbf{1.209}& 1.266& 1.228& 1.165& 1.207&1.358\\
 \cline{2-10}
 &\multirow{3}{*}{1} & 0& 1.191& \textbf{1.175}& 1.205& 1.197& 1.184& 1.202&1.775\\
 & & 10& 1.193& \textbf{1.200}& 1.218& 1.193& 1.153& 1.185&1.405\\
 & & 20& 1.143& \textbf{1.191}& 1.269& 1.223& 1.096& 1.219&1.366\\
 \cline{2-10}
 &\multirow{3}{*}{10} & 0& 1.196& \textbf{1.183}& 1.211& 1.195& 1.228& 1.198&1.863\\
 & & 10& 1.193& \textbf{1.181}& 1.270& 1.204& 1.145& 1.181&1.105\\
 & & 20& 0.910& \textbf{1.179}& 1.268& 0.931& 1.023& 1.266&0.994\\
 \cline{1-10}
 \multirow{9}{*}{1.8}&\multirow{3}{*}{0.1} & 0& 1.808& \textbf{1.790}& 1.823& 1.808& 1.826& 1.804&1.537\\
 & & 10& 1.743& \textbf{1.804}& 1.774& 1.685& 1.743& 1.816&1.285\\
 & & 20& 1.302& \textbf{1.787}& 1.499& 0.894& 1.307& 1.484&1.241\\
 \cline{2-10}
 &\multirow{3}{*}{1} & 0& 1.805& \textbf{1.793}& 1.793& 1.788& 1.819& 1.780&1.567\\
 & & 10& 1.645& \textbf{1.798}& 1.721& 1.654& 1.632& 1.725&1.305\\
 & & 20& 0& \textbf{1.779}& 1.312& 0.780& 1.458& 1.309&1.125\\
 \cline{2-10}
 &\multirow{3}{*}{10} & 0& 1.783& \textbf{1.822}& 1.722& 1.702& 1.756& 1.711&1.776\\
 & & 10& 1.297& \textbf{1.833}& 1.557& 1.196& 1.305& 1.568&1.269\\
 & & 20& 0& \textbf{1.786}& 0.899& 0.019& 0.508& 0.938&0.706\\
\bottomrule \label{methods}
\end{tabular}
}
\end{center}
\end{table}
\linespread{1}

\vspace{-0.5cm}
\linespread{1}
\begin{table}[htbp]
\begin{center}
\caption{Running time (s) of different algorithms in various scenarios}
\scalebox{0.8}{
\begin{tabular}{cccccccc}
\toprule
Parameter pair&None& U-net++&SS&Kalman&EMD&LMP&Clipping\\
\midrule
(1.2,0.1,0)&1.78 &\textbf{17.52} & 2.79& 3.24& 17.83& 127.45& 3.02\\
(1.2,1,10)&1.84 &\textbf{17.37} & 2.36& 3.16& 15.19& 156.28& 3.75\\
(1.2,10,20)&1.81 &\textbf{17.63} & 2.54& 3.28& 15.68& 198.03& 3.44\\
(1.8,0.1,0)&1.86 &\textbf{18.02} & 2.21& 3.46& 14.22& 768.14& 3.62\\
(1.8,1,10)&1.90&\textbf{17.12}&2.32&3.09&14.6&1127.07&3.64\\
(1.8,10,20)&1.84 &\textbf{16.98} & 2.46& 3.17& 14.17& 3688.05& 3.39\\
\bottomrule \label{time}
\end{tabular}
}
\end{center}
\end{table}
\linespread{1}
\vspace{-0.4cm}

From Table \ref{methods}, we can see that the proposed U-net++ based algorithm can effectively estimate noise parameter under all considered environments. It performs much better than these baselines, especially for the scenario where the IN component is not obvious. This is because these baselines can not well alleviate the negative effect of transmitting signal on noise parameter estimation based on single-channel received signal. Meanwhile, the performance gain of our proposed U-net++ is achieved at the cost of relatively longer running time. Nevertheless, the running time of our method is still acceptable compared with that of EMD and LMP, as it is shown in Table \ref{time}.
\vspace{-0.3cm}

\section{Conclusion}
In this paper, we address the issue of mixed noise parameters estimation based on received single-channel signal including both transmitting signal and noise. To mitigate the negative influence of transmitting signal, we propose an U-net++ based framework to separate transmitting signal and mixed noise. We further utilize clipping preprocessing and power normalization to suppress the outliers due to impulsive noise. Extensive simulation results show that our method can effectively estimate mixed noise parameters under different scenarios. Moreover, it can achieve better performance compared to exiting blind source separation based methods, especially when GSNR is relatively large and impulsivity is not obvious.

\footnotesize
\bibliographystyle{IEEEtran}
\bibliography{ref}

\end{document}